\documentclass[preprint,12pt]{elsarticle}

\usepackage{amssymb}
\usepackage{amsmath}

\journal{Physica A}

\begin{document}

\begin{frontmatter}

\title{Interests Diffusion in Social Networks}

\author[1,3]{Gregorio D'Agostino} 
\author[2]{Fulvio D'Antonio} 
\author[1,4]{Antonio De Nicola} 
\author[4]{Salvatore Tucci} 

\address[1]{ENEA: Italian National Agency for New Technologies, Energy and Sustainable Economic Development, Rome, Italy}
\address[2]{Universit\`a Politecnica delle Marche}
\address[3]{Center for Polymer Studies, Boston University, Boston, Massachusetts, United States of America}
\address[4]{University of Tor Vergata, Rome, Italy}

\begin{abstract}
Understanding cultural phenomena on Social Networks (SNs) and exploiting the implicit knowledge about their members is attracting the interest of different research communities both from the academic and the business side. The community of complexity science is devoting significant efforts to define laws, models, and theories, which, based on acquired knowledge, are able to predict future observations (e.g. success of a product). In the mean time, the semantic web community aims at engineering a new generation of advanced services by defining constructs, models and methods, adding a semantic layer to SNs. In this context, a leapfrog is expected to come from a hybrid approach merging the disciplines above. Along this line, this work focuses on the propagation of individual interests in social networks. The proposed framework consists of the following main components: a method to gather information about the members of the social networks; methods to perform some semantic analysis of the Domain of Interest; a procedure to infer members' interests;
 and an interests evolution theory to predict how the interests propagate in the network. As a result, one achieves an analytic tool to measure individual features, such as members' susceptibilities and authorities. Although the approach applies to any type of social network, here it is has been tested against the computer science research community.
 The DBLP (Digital Bibliography and Library Project) database has been elected as test-case since it provides the most comprehensive list of scientific production in this field.

\end{abstract}


\end{frontmatter}

\section*{Introduction}

Social networking platforms (SNPs) collect a huge amount of information (and hence implicit knowledge) about their members and different domains of interest. Such knowledge concerns interests, friends, best practices, activities and other facts of life. General purpose SNPs (e.g., Facebook, Twitter) contain information on several domains (e.g., music, movies, literature, travels) whereas domain-specific SNPs, e.g., anobii and LinkedIn collect information on specific topics (e.g., books for anobii and job and careers for LinkedIn).

According to McKinsey industry report of 2011 \cite{Manyika2011}, the total volume of worldwide dispersed data is increasing at a rate of about 50\% per year, that is around a 40-times growth in ten years. 
Data storage is becoming almost free since hardware devices are almost inexpensive and SNP companies gain added value by gathering data about members \cite{Dhar:2013}. 
Management and analysis of big data involved in social networks are among the most effective activities in the scope of Data Science  \cite{Dhar:2013}.

Organizing and managing information conveyed by SNPs in order to extract knowledge about their members is leading the market to a new generation of services focused on specific users' needs.
 There is a big opportunity for a paradigm shift from decisions based on "gut feelings" to decisions based on data analysis. 
 Advanced applications exploiting social network knowledge can generate value in different sectors, such as, security, politics, business, and "social good" \cite{akerkar_big_2013}.

In this paper we study temporal evolution of people's interests in social networks. The objective is to understand basic mechanisms and to estimate some individual features such as susceptibility and authority; that is measuring the tendency of a person to be influenced by her/his connections and her/his tendency to influence others. Estimation of these human characteristics can be used, in the long run, as a basis for the development of advanced marketing services targeted to specific individuals. 

A \textit{social network} (SN) consists of a community of "members" linked together with some kind of relationships (e.g., friendship, coauthorship, co-working). 
A SN is a virtual artifact originated from human activities. Developing a service leveraging on SN knowledge requires a hybrid approach based on both engineering and natural science techniques and methodologies \cite{Hevner:2004}. From this perspective, we may study the temporal evolution of people's interests as a dynamic phenomenon arising in an anthropic system. 
Our hypothesis is that this phenomenon results from the combined action of several factors: people connections, general trends, pre-existing interests and both the attitudes of people to be influenced by or to influence others. Furthermore, we deem that, given an application domain, temporal evolution of interests depends on the topics, since people can be susceptible to some specific information more than to others: 
e.g., American people are usually interested on the Super Bowl rather than Europeans; whereas it is the other way around for the final of the Champions League.   

The interest propagation phenomenon in social networks has been already studied by different disciplines \cite{vespignani2012modelling} through different approaches: 
data mining, complexity science, semantic, and social science.  

In \cite{Domingos:2001} \cite{Richardson:2002}, the authors propose a data mining approach to estimate the propagation of events (e.g. threads) and the identification of influential members. 
Most of the efforts in the data mining community have been devoted to define progressive models. In such models, once a node (member) becomes active (interested in a topic), it remains active. 
The most important propagation models are the Independent Cascade Model (ICM) and the Linear Threshold Model (LTM). 
Both of the previous models were first introduced in \cite{Kempe:2003}. The key characteristic of ICM is that diffusion events along every arc in the social network 
are mutually independent; while the key characteristic of LTM is that members change their behaviour if they are exposed to multiple independent sources.
Another data mining approach was presented in \cite{Goyal:2010}. Here the authors propose models and algorithms to learn influence probabilities parameters from a "social graph" and a log of actions by the users.

Complexity science includes the study of complex networks \cite{Barab‡si15101999} \cite{watts}. Among the phenomena treated by this discipline, epidemics \cite{PhysRevLett.86.3200, vespignani2012modelling} studies 
the spread of viral processes in networks. The complexity science is mainly focusing on human infectious diseases and software malware spread. 
However there is a growing interest in studying topics diffusion in social networks \cite{Wang:2011:ISC:1963405.1963508},
 social dynamics\cite{castellano2009statistical, galam2005local} or even non  consensus dynamics \cite{shao2009dynamic}.

Merging the topological and semantic analysis of social networks represents a new and potentially fruitful research field which is providing promising results \cite{Jung:2007:TSS:1419662.1419688} \cite{mika2007ontologies} \cite{Boj?rs200821} \cite{Kinsella2009121} \cite{Cucchiarelli2012}. Our work shares the use of a semantic conceptual representation of a Domain of Interest 
\cite{Poggi:2008:LDO:1793934.1793939} in the social network context with the formers.

A social science approach is presented in \cite{Aral20072012}. There, the authors describe an experiment performed on Facebook to estimate influential and susceptible members of social networks with respect to some social features, such as age and sex. Another interesting issue considered by the social science community is homophily 
(i.e., the tendency for individuals to choose friends with similar tastes and preferences) \cite{manski1995identification} \cite{aral2009distinguishing}. Our work does not deal with such issues. 

In the context of social science, the concept of "meme" \cite{dawkins2006selfish} is acquiring a growing attention representing the elementary brick for the evolution of culture and behavior in the human communities. As such the meme is different from our concept of interest; however some authors \cite{adar2005tracking} have employed the term as synonym of our concept of interest.

We treat the SN as a physical system and we model interests dynamics as a diffusion process. Like a physical system, a thermodynamic equilibrium is reached after a certain time period when no heat source is applied. 
Similarly in SNs, arising of new topics can be considered as a heat source that hinders the equilibrium of interests thus preventing all people to be interested in the same topics.

Our approach is based on the analysis of social network's connections and the temporal evolution of the interests of its members. We have defined a general Markov evolution process and we have tested it on a co-authorships network in computer science. Although our paradigm is very general, we have deeply analyzed the DBLP\footnote{DBLP: Digital Bibliography \& Library Project. http://www.informatik.uni-trier.de/~ley/db/} computer science bibliography that provides an exhaustive list of papers in computer science from the onset of the discipline to present. In this application, we infer people interests from the titles of the documents they authored by means of natural language processing (NLP) \cite{Navigli:2004:LDO:1105710.1105712}.

The main contribution of this work is a framework consisting of four main building blocks:
\begin{itemize}
\item A modelling approach for social networks, to give an explicit specification of SN knowledge concerning people, their relationships and their interests.
\item A diffusion theory, to describe the interest propagation phenomena and to make predictions about them.
\item A method to measure individual features (i.e., people susceptibility and authority).
\item A software application, to assess the theory and to measure individual features.
\end{itemize}

In the following the above-mentioned building blocks are presented along with the outcomes of the analysis of the DBLP dataset. In particular, Section II presents the modelling approach to represent the implicit knowledge of social networks. Section III describes the interest propagation theory. Section IV presents the case study and Section V describes some validations of the theory. Finally, Section VI discusses the findings for the case study.

\section*{Knowledge Representation of Social Networks}

In this Section we present our approach to unveil tacit knowledge encompassed in social networks and to turn it into explicit and formal knowledge \cite{nonaka1998concept}. 

We start from an abstract representation of a SN that results from the set of relationships among the members of the real social network. The second problem is to provide a reliable representation of the semantics of the Domain of Interest \cite{Poggi:2008:LDO:1793934.1793939}. Finally, we need a means to represent connections between these two systems. The former elements provide an abstract representation of the system we intend to deal with at a given time. However the knowledge contained in the real system is not limited to the instantaneous configuration but it results also from the chronology of the events. 

\begin{figure}[!ht]
\begin{center}
\includegraphics[width=0.80\textwidth, angle=0]{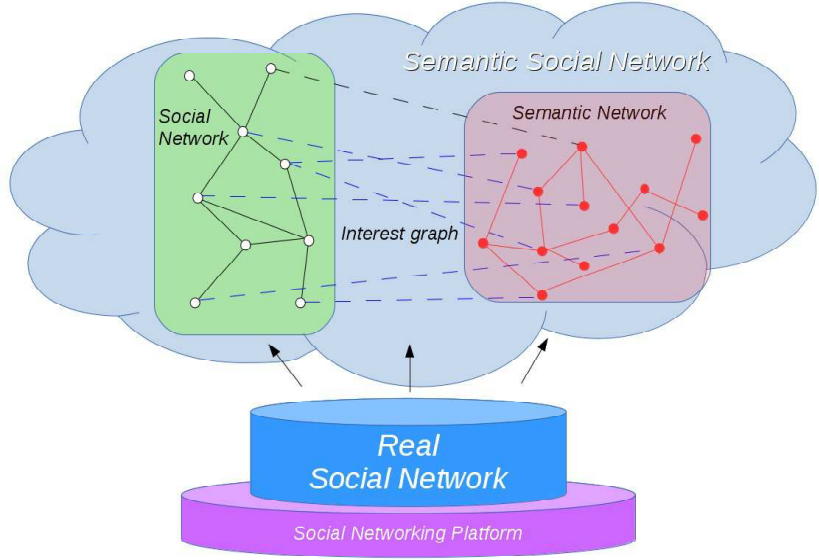}
\end{center}
\caption{A pictorial representation of our approach to semantic social networks. The social networking platform, at the bottom, supports the activities of a real social network. 
The latter is represented by the semantic social network (the cloud) consisting of the social network (at left), a semantic network (at right), and the weighted links from the interest graph.}
\label{GeneralRepresentation}
\end{figure}

As anticipated in the Introduction and shown in pictorially represented in Fig. \ref{GeneralRepresentation}, social networking platforms (SNPs) support the activities of a real social network (SN). 
The latter can be represented as a semantic social network (SSN) that, in turn, consists of a social network (SN), a semantic network (SeN) \cite{sowa2006semantic}, and a weighted interest graph (WIG) connecting them.  

\subsection*{Interests in Social networks}

A SN can be represented by a directed graph $SoN = (H,F)$, where the set  $H$ of nodes $\{h_i\}$ represents the members of the social community $H= \{ h_1, h_2, ..., h_{|H|}\}$ and the set $F$ of links $f_{i,k}$ represents relationships between members as ordered pairs $F=\{ f_{1,1}, f_{1,2}, ..., f_{|F|}\}$.

Expressions of interest are events (e.g., filling or changing a form profile, posting, commenting or "liking" a post, publishing a paper) demonstrating a positive attention by a member to a product. All possible products form the Domain of Interest (DoI). It is worth mentioning that the term product here is employed in its broad sense, referring not only to goods, but also to cultural events and scientific products such as articles, books, movies, etc. 

\subsection*{The Semantic Network}

Conceptual images of products can be expressed in terms of a finite number of concepts belonging to a semantic network representing the DoI. 
A semantic network can be seen as a graph $ SeN =(\Lambda,R) $ where the set $\Lambda = \{ \lambda_1, \lambda_2, ..., \lambda_{|\Lambda|}\}$ 
of nodes are concepts (logos) and $R= \{ r_1, r_2, ..., r_{|R|}\}$ are the links that represent  semantic relationships of different types as subsumption, 
meronimy and similarity \cite{Resnik:1999} between the different concepts. 

Given the semantics structure, we further assume that there exists a set of elementary concepts, that we name "topics" $C= \{ c_1, c_2, ..., c_{|C|}\}$, 
such that one can associate to each product (or its abstraction) a subset of topics. 
The identification of basic topics plays a fundamental role and is a critical issue treated by the ontology engineering discipline \cite{DeNicola2009258, debruyne2013grounding, Ferndndez1997, suarez2012ontology, Sure2003} 
involving both automatic techniques (such as natural language processing) and domain experts' validation. We further discuss this point in the test case. 

\subsection*{The Semantic Profiling Process}

A \textit{Interest Graph} ($IG$ see Fig. \ref{GeneralRepresentation}) represents an abstraction of a community of people together with their interests. It can be represented as a bipartite graph consisting of a set of nodes $N$ partitioned in two groups, one representing people and the other set, $C= \{ c_1, c_2, ..., c_{|C|}\}$, representing the topics, and a set of relationships $I$ representing the interests of people in topics. Consequently, $IG=(H,C,I)$, where $I= \{ i_1, i_2, ..., i_{|I|}\}$, and  

\begin{equation}
\label{IeN}
i_i = (h_j, c_k)  \text{ with } h_j \in H  \text{ and } c_k \in C.
\end{equation}

A \textit{Weighted Interest Graph} $WIG$ is an $IG$ with weights assigned to the links between people and topics. Such links can represent, for instance, either the probability to be interested or the degree of interest in a topic. Consequently, $WIG =(P,C,I,W(I))$, where $w(I)$ is a mapping from the set of relationships $I$ to the [0, 1] range.
 
\begin{equation}
\label{WIG}
w(I): I \rightarrow  [0, 1]
\end{equation}

\begin{figure}[!htb]
\begin{center}
\includegraphics[width=1.00\textwidth, angle=0]{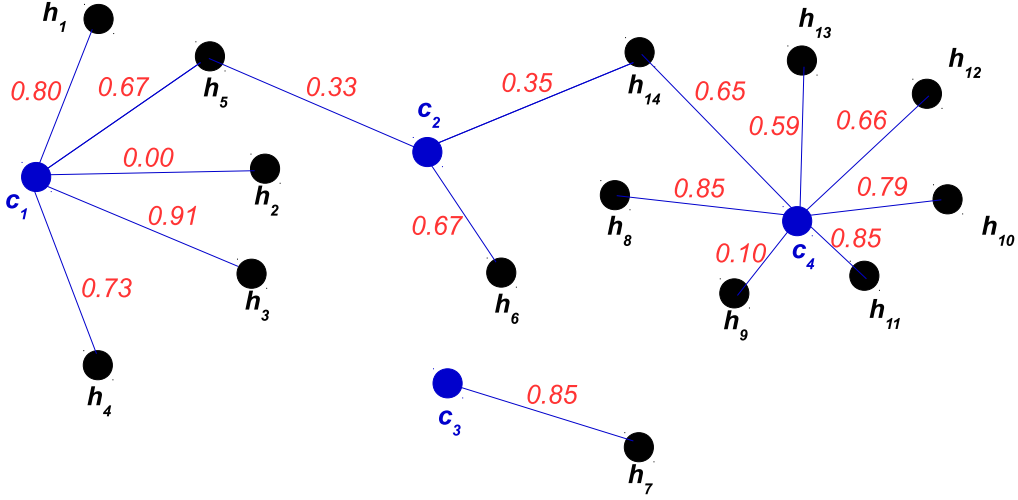}
\end{center}
\caption{A representation of a weighted interest graph as a bipartite graph}
\label{WeightedInterestGraph}
\end{figure}

\noindent Fig. \ref{WeightedInterestGraph} shows a representation of a weighted interest graph as a weighted bipartite graph.

We define \textit{semantic profiling} the process of associating  interests to the members of the SN, that is inferring links and their relative weights of the WIG.
The set of interests characterizing a member $h_i$ is defined as her/his \textit{semantic profile} $S_{h_i}$:

\begin{equation}
\label{SemProf}
S_{h_i} =\{ w_k:(h_i,c_k) \in I\} 
\end{equation}

\noindent where  $c_k\in C$,  $k\in(0,|C|)$, $h_i\in H$  and $w_k = w(( h_i, c_k))$.

\noindent In other words the \textit{semantic profile} of a member is a subset of the \textit{interest graph}.

Given the basic set of interests $c_k$, one possible choice to provide a member $h_i$ with a semantic profile is to attribute a likelihood $L_{h_i}(c_k)$. 

The \textit{semantic profile} of each member is not static and one needs to account also for its temporal evolution: 

\begin{equation}
\label{SemProfDynamics}
S_{h_i}(t) =\{ c_k:(h_i,c_k) \in {I(t)} \} 
\end{equation}

\noindent where  $c_k\in C(t)$,  $k\in(0,|C|)$, $h_i\in H(t)$  and $w_k(t)= w(( h_i, c_k))$.

\noindent By introducing the scalar product 

\begin{equation}
\label{scalar} 
\langle w_{h_i} \cdot w_{h_j} \rangle =  \sum_{c_k} w_{h_i}(c_k) w_{h_j} (c_k), 
\end{equation} %
 
\noindent One may treat the \textit{space of semantic profiles} as an euclidean space.  One can, also, define the quadratic distance between two semantic profiles:
 $ d^2(w_{h_i} , w_{h_j}) = \langle (w_{h_i} - w_{h_j})^2 \rangle  $. The smaller is the distance the closer are the members' interests.
 
 \subsection*{The Semantic Social Network}
 A \textit{semantic social network} \textit{SSN} represents the relationships between members, the semantics of the Domain of Interest and the actual interests of the community of members with their weights. 
 From the mathematical point of view it can be formally written as a set of six entities: $SSN = (H, F, \Lambda, R, I, W_I) $
 where \textit{H} represents the set of members (humans); \textit{F} represents the relationships between members; 
 $\Lambda$ represents the set of concepts; \textit(R) represents the set of semantic relationships; \textit{I} represents the interest of people on topics; 
 and $W_I$ represents the degree of interests in topics of people.
 
Semantic social networks are like living entities: they are born, grow, shrink and, finally, die (close). Appearance of new nodes may represent both inclusion of new members and emergence of novel topics. Similarly, disappearances of nodes may mimic the cease of participation of people to the community or the obsolescence of topics. Moreover interests of members on topics may change their intensity during the time.
To model the dynamics of the latters, we define a \textit{dynamic semantic social network}
$ SSN_t = (H(t), F(t), \Lambda(t), R(t), I(t), W_I(t))  $

\section*{Interest Propagation Dynamics}
In this Section we assess a model of \textit{interest propagation} to predict the evolution of the interests in a \textit{semantic social network}. It accounts for the structure of the social network and its evolution \cite{palla2009statistical} without predicting it. Consequently, it has the objective to estimate the probability for a person $h_i$ to be interested in a topic $c_k$ at a given time.

The resulting \textit{equations of dynamics} are based on the following four assumptions:

\begin{itemize}
\item As a person, each member tends to keep her/his own beliefs.
\item Each member is partly influenced by others interacting with her/him (one to one interaction).
\item Each member is partly influenced by trends (one to all interaction).
\item The evolution mechanism is markovian.
\end{itemize}

The evolution equations resulting from the above assumptions can be approximated for short time increments:

\begin{equation}
\label{InterestPropagationEquation}
L_{h_i} (c_k, t+\Delta t) = \left[1-x_i(c_k)-\tilde{x}_{si}(c_k)\right] \cdot L_{h_i} (c_k, t)+\frac 1{|N_{h_i}|}\cdot \sum_{h_j \in N_{h_i}} x_{ij}(c_k) \cdot L_{h_j}(c_k, t)+x_{is}(c_k) \cdot L_s(c_k, t)
\end{equation}

\noindent The three addendums at the right hand side, respectively, model the personal tendency of a person to keep interest in a topic $c_k$, the influence of the neighbours and that of the environment.
 In particular,  $L_{h_i} (c_k, t+ \Delta t)$ represents the probability of person $h_i$ to be interested in the topic $c_k$ at time $t+\Delta t$. $L_{h_i} (c_k, t)$ represents the probability of person $h_i$ to be interested in the topic $c_k$ at time t.  $L_s (c_k, t)$ is the probability for the environment to provide some information on interest $c_k$ at time $t$. We refer to this quantity as the "source term".  $x_{i}(c_k)$ and $x_{ij}(c_k)$ are parameters (to be experimentally determined) that characterise the different individuals.
We do assume that when all neighbours share the same interests (i.e. their profiles) the interest profile should not experience any variation, therefore:

\begin{equation}
\label{xi}
x_i(c_k) = \frac{1} {|N_{h_i}|} \sum_{h_j \in N_{h_i}}   {x_{ij}(c_k)}
\end{equation}

\noindent and similarly, when the single member profile equals the trends source, no influence is expected; that is $\tilde{x}_{is}=x_{is}$.

In the limit for $\Delta t \to 0$ the above discrete-time equations tend to heat-like equations:

\begin{equation}
\label{ContinuousInterestPropagation}
\frac{\partial}{\partial t}  L_{h_i} (c_k, t) = \left[-v_i(c_k)-v_{si}(c_k)\right] \cdot L_{h_i} (c_k, t)+\frac 1{|N_{h_i}|}\cdot \sum_{h_j \in N_{h_i}} v_{ij}(c_k) \cdot L_{h_j}(c_k, t)+v_{si}(c_k) \cdot L_s(c_k, t)
\end{equation}

\noindent where $v_{ij}(c_k) = \lim_{\Delta t \to 0} x_{ij}(c_k)/ \Delta t $ represent the rates of susceptibility per unit time. It should be noted that $x_{ij}(c_k)$ depend on time increment; such a dependence has not been made explicit for the sake of brevity.

The evolution equations (\ref{InterestPropagationEquation}) do not lead to consensus, that is, the L's do not converge to a common value. However, the system becomes ergodic (in all its connected parts) when the susceptibilities from the environment are removed. In fact, there exists a suitable weighted average of the L's that is a conserved quantity: 

\begin{equation}
\label{conserved}
\hat{L}(c_k) \stackrel{def}{=} \sum_{h_i} L_{h_i}(c_k, t) \cdot b_i(c_k);
\end{equation}

\noindent where the constant $b_i$ satisfies the equilibrium equations:

\begin{equation}
\label{secular}
\sum_{h_j} \frac{1} {|N_{h_j}|} x_{ji}(c_k) \cdot b_j(c_k) = b_i(c_k) \cdot x_i(c_k).
\end{equation}

\noindent When the system is isolated all the L's tend to $\hat{L}$. However there are many reasons for which this condition is never reached. In fact the topology of the network is dynamic, the susceptibilities may also change during the time and the environmental influence is not negligible.

When all the mutual susceptibilities $x_{ij}$ are equal, the conserved quantity of the eq. (\ref{conserved}) acquires a pure topological form: 

\begin{equation}
\label{conserved2}
\hat{L}(c_k) = \frac {\sum_{h_i} L_{h_i}(c_k, t) \cdot N_{h_i}} {\sum_{h_i} N_{h_i}} ;
\end{equation}

\noindent  that is, authors influence the asymptotic "consensus profile" according to  their degrees.

Key concepts in the interest propagation theory are the individual features, i.e., \textit{susceptibility} and \textit{authority}, characterizing a person with respect to a specific Domain of Interest.

According to Merriam-Webster\footnote{http://www.merriam-webster.com/}, \textit{susceptibility} is defined as the "state of being easily affected, influenced, or harmed by something". Here, in particular, there are three different parameters related to it: $x_{ij}(c_k)$ and $x_{is}(c_k)$.

 $x_{ij}(c_k)$ is a positive number representing the attitude of a member $h_i$ to be influenced by each of her or his neighbours $h_j$ with respect to the topic $c_k$. 
 
 The $x_i$ parameter measures the susceptibility of a member $h_i$ to her/his neighbours' total solicitation with respect to the topic $c_k$. It is given by the average of $x_{ij}$ over all $j$'s (as in eq. \ref{xi}).

Finally, $x_{is}(c_k)$ represents the attitude of a member to be influenced by the general trends (i.e., environment or \textit{trends susceptibility}).

According to Merriam-Webster, the authority is the "power to influence or command thought, opinion, or behavior".
We may introduce $a_i$ that measures individual authority as following:

\begin{equation}
\label{ai}
a_i \stackrel{def}{=} \sum_{h_j \in N_{h_i}}   {x_{ji}(c_k)} 
\end{equation}

It is worth stressing that the $a_i$'s do measure a sort of "local" authority as the capability to influence the whole systems depends on the topology of the social network and its changes during time. The $b_i$'s of Eq. \ref{secular} may represent a sort of global authority if the social network were static. 

\section*{Interests Dynamics in the Research Community}

One of the possible applications of our general framework is analysis of publications in the research community. In this respect, 
according to our knowledge representation approach, we identify the members of the social network 
with the authors connected by the co-author relationships (representing the edges). 
In principle other types of relationships, such as belonging to the same institution, participation to common projects and be accounted for a paper should be considered. 
However to handle such variety of links, one should resort to multiplex approach that has been intensively investigating during the last decade \cite{quattrociocchi2014opinion,gomez2013diffusion} and is beyond the scope of this work.

In principle, all the information contained in the whole set of communications (papers, talks, interviews) form the corpus where to extract the semantic network. However access to all this information is impossible and, hence, an approximate conceptualization of the Domain of Interest is necessary. 

The expressions of interest can be of different types. The most significant is the publication of new scientific products (e.g., paper, book), but there are others such as the citation of a work, the invitations to conferences, attendance to talks, seminars, conferences and other presentations, etc. All these events contribute to the semantic profiling of a member.  

One of the most difficult tasks for researchers in the field of social network analysis is to obtain a significant dataset to test new methods and software. In fact, despite the hardware to store information being inexpensive, there are privacy issues and business motivations that hinder the process of making these datasets open to the research community. Open data \cite{auer2007dbpedia} provide a means to oppose this tendency. Their availability is crucial for the advancement of research in data science. Moreover, since we are interested in temporal evolution, datasets need also to carry the chronological information, which is even harder to attain.

Fortunately, existing repositories of information about scientific research papers provide free and open source datasets. 
In particular, the DBLP dataset provides a comprehensive list of scientific production in computer science. 

\subsection*{Computer Science Case Study}

The goal of the experimental evaluation is to test the theory (that is eq. \ref{InterestPropagationEquation}) against a real case study represented by the computer science community.
In order to perform the analysis, we need to acquire the information about the topics defining the scope of the computer science domain and the evolution dynamics of both the social relationships and the interests of the authors.
In principle, the above information could be extracted from different sources, however the DBLP dataset provides both the information through a single XML document [35]. 

The analysis of the test case consists of the following steps:
\begin{itemize}
\item{\textit{papers selection}, according to their type (e.g., journal, conference, book chapter) and year;} 
\item{\textit{interests and topics identification};}
\item{\textit{papers indexing};}
\item{\textit{identification of social network topology and its temporal evolution};}
\item{\textit{semantic profiling of the members}};
\item{\textit{analysis of the trends};}
\item{\textit{assessment of model parameters}.}
\end{itemize}

\subsubsection*{Papers Selection} \label{treatable}
The DPLP database is an evolving entity. Results presented in this work refer to the dataset as published by november 2013, it consists of 2.360.780 papers and 1.337.857 authors. The observation period has been limited to years from 1950 to 2012. In such a temporal range, the number of considered papers is 2.246.098 and the authors 1.337.195. In order to study the evolution of authors' interests it is necessary to observe some change in their semantic profile during time; therefore only authors that have published papers in, at least, two different years can be analysed. We have named those authors "treatable". It is worth noting that only 519.886 authors out of 1.337.195 are treatable as far as 2012. It is reasonable to image that this is mainly due to students that just publish one work and then leave the world of research for other activities. It should be noted that not-treatable authors are intrinsically untreatable, i.e., independently from the specific capability of a suitable set of topics to index papers.

\subsubsection*{Identification of Interests and Topics}

As said all the communications form the potential corpus for the reconstruction of semantic network. However we have limited our analysis to the titles. There are other possible choices such as the abstract and or the introduction; however, it is worth noting that a lot of information contained in the papers (and specifically in those sections) do not refer to their specific contents, but to the general state of the art in the field and, hence, the semantic analysis of the full text (or the abstract and introduction) could include spurious terms not related to the subject.  Moreover, introductions were not available for all papers. Finally the analysis of millions of full papers is extremely time consuming and may be not sustainable. For all the  reasons above, limiting the analysis to the titles seems appropriate.

We have used natural language processing techniques \cite{Navigli:2004:LDO:1105710.1105712} 
to extract multi-lexemes from the corpus of titles. 
Then we have validated them by human processing devoted to remove general purpose lexemes 
that are not specific of the computer science domain and to merge synonyms. 
The resulting list of lexemes forms the set of basic topics  $\{c_k\}$, that we have used for the analysis. 
Recently, based on Latent Dirichlet Allocation, new methods have been employed 
for automated topic extraction \cite{lancichinetti2014high}.
These novel techniques will possibly improve the quality of the set.

Figure \ref{TopicEvolution} shows the evolution of the number of detected topics within a 
given year ($n_c(t)$). In the range 1950-2012, 7.632 topics have been 
identified by using the TermExtractor web application \cite{sclano2007termextractor}. 
This is a tool that allows extracting the shared terminology of a community from 
the available documents in a given domain. Figure \ref{TopicEvolution} clearly shows that the field has experienced an exponential proliferation of concepts and reached its maturity at the beginning of the XXI century. This is consistent with the common perception of the addicts.

We have mentioned that the selection of the basic set of topics $C=\{c_1, c_2, \ldots, c_N\}$ plays a crucial role to "tame" the Domain of Interest. 
It is worth stating that, in order the diffusion theory to work,  the $c's$ must form a "basis" for the algebra of interests. The most relevant relations among concepts (and hence among interests) are generalization (specialization) and similarity. From the algebraic point of view these represent inclusion relationships.  The two constraints we impose on the set $C$ are "completeness" and "independence" respectively. 

A set of concepts will be named "complete" when each concept can be seen as the union of a subset of basic concepts:

\begin{equation}
\label{decomposition}
\forall c \exists \{ i_1, i_2, \ldots, i_m\} : c=\cup_{k=1}^m c_{i_k}.
\end{equation}
\noindent this means that each interests is the combination of a set of the basic interests. 

On the other side a set of concepts will be named "independent" when each pair is disjoint, that is there does not exist a concept representing a common specialization of both:

\begin{equation}
\label{independency}
\forall c_i, c_j  : \ c_{i} \cap c_j =\emptyset.
\end{equation}
\noindent When eq. (\ref{independency}) holds the decomposition of eq. (\ref{decomposition}) is unique.

We identify the basic topics with a subset of the multi-lexems. The quality of the results do strongly depend on the capability of the selected set of multi-lexemes to fulfill the required constraints.

The former analysis refers to topics as autonomous entities, however they belong to a semantic network. The analysis of semantic structure of the Domain of Interest is beyond the objectives of the present work.
 
\subsubsection*{Papers Indexing}

Once the set of topics is assessed, it is possible to attribute a subset of them to each scientific product. Conversely, each topic can be given a frequency as the number of papers referring to it ($\nu (c_k,t)$).

\begin{equation}
\label{ExpValueTopicFrequencyEq}
\nu(t)= \frac{\sum_{c_k} \nu (c_k,t)} { n_c(t)} 
\end{equation}

where $\nu(c_k,t)$ is the number of the occurrences of the topic $c_k$ and $n_c(t)$ is the number of the topics up to a given year.

\begin{figure}[!ht]
\begin{center}
\includegraphics[width=0.70\textwidth, angle=0]{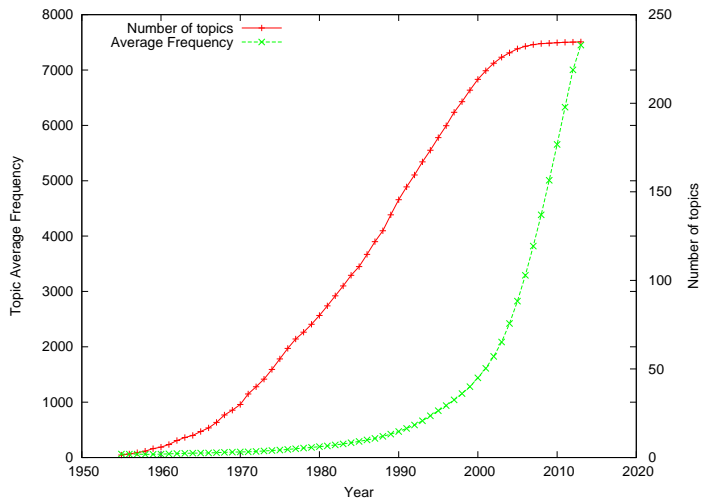}
\end{center}
\caption{Temporal evolution of the number of topics and their average frequency of publication as 
resulting from the analysis of the DBLP database by natural language processing.}
\label{TopicEvolution}
\end{figure}

Figure \ref{TopicEvolution} shows the evolution of the average frequency of the topics during the time.
Even if the semantic complexity of the 
field reached its mature state (no further topic proliferation), the production in the field is still growing.

\subsubsection*{Topology and Evolution of the Social Network}
We have focused our research on the time period from 1950 to 2012. For each year, the nodes of the social network are given by the authors that have written papers by that year and the edges are given by the co-authorships. According to this assumption, the social network evolves incrementally with the time. At a given a year, we attribute a link to all authors that have published a product together by that time. We never remove links and, therefore, our social network shows increasing complexity. It is worth noting that in real life authors stop to publish (e.g., for retirement or job change) or give up their collaboration. However we have not taken into account such phenomena since we are limiting our work to the information available in the DBLP dataset. Moreover, all members are treated on the same ground regardless of their notoriety or scientific production. The former approximation may incide on the quality of our results.  

Consistently with Barabasi-Albert \cite{Barab‡si15101999}, 
the observed distribution of authors' collaborators is a power law with 
an exponent oscillating within the [2.5, 3.0] range. 

\subsubsection*{Semantic Profiling}

For each year in the observation period, a semantic profile can be given to each author by means of the relative frequencies of "expressions of interest" (publications):

\begin{equation}
\label{ShareOfInterest}
\xi_{h_i}(c_k,t) = \frac{\nu_{h_i}(c_k,t)}{ \sum_{c_k} \nu_{h_i}(c_k,t) },
\end{equation}

\noindent where $\nu_{h_i}(c_k,t)$ represents how many papers, written before the considered year, are indexed by the topic $c_k$. This function by definition spans the $[0,1]$ range; the unitary value represents a total interest in the subject while a null value means no interest at all. These semantic profiles represent our estimates of the probabilities $L_{h_i}(c_k,t)$ evolving according to eq. (\ref{InterestPropagationEquation}) that is, one can estimate the likelihood of an author $h_i$ to publish on a topic $c_k$ through its share of interest:

\begin{equation}
\label{InterestEstimation0}
L_{h_i} (c_k,t) \sim \xi_{h_i}(c_k, t)
\end{equation}

Treatable authors (as defined in \ref{treatable}) are named "semantically treatable" once they acquire a not-trivial semantic profile. Their number is affected by our capability to identify interests related to the domain and to index all papers. In principle, we are able to estimate both trends and neighbours susceptibilities of semantically treatable authors.  However, the higher is the number of publications of an author the more precise is the estimation of susceptibility.

Sewing together all the semantic profiles of the different members one achieves  the interest graph.

To reconstruct the time evolution of the social network, for each year, we have attributed a link to all pairs of members sharing a paper published by that year. We have not attributed weights to links depending neither on the age of the shared publications nor on their number or scientific relevance. The former are two strong hypotheses that may be relaxed in future (ongoing) works. Human contacts that took place in a remote past may have ceased; moreover, a successful publication  may stimulate further scientific common activity more than a coarse one. 

\subsubsection*{Analysis of the Trends}

The analysis of trends in social networks has been exploited in different cases including the prediction of stock market \cite{Curme12082014}. Generally speaking one needs a source to evaluate statistics of different topics. In our case the popularity of a topic can be estimated by its relative frequency over all published papers: 

\begin{equation}
\label{TopicsProbEq}
\xi_s(c_k,t)= \frac{\nu(c_k,t)} {\sum_{c_k} \nu(c_k,t)} 
\end{equation}

\noindent where $\nu (c_k,t)$ is the frequency of the topic $c_k$ at time $t$.

\noindent It can be regarded as the likelihood of a random person to be interested in the concept ${c_k}$ at time $t$.

As can be seen from Figure \ref{TopicEvolution}, the number of topics has been saturated during the last ten years. The same thing applies to the relative relevance of the topics. 

\noindent One can estimate the entropy of the Domain of Interest by means of the Shannon Entropy of the topic frequencies: $H(t) = - \sum_{c_k} \xi (c_k,t) \cdot ln[\xi(c_k,t)].$ 

\noindent It provides an index of dispersion for the relevance of the topics. Also entropy experiences  saturation for larger time. %

\section*{Assessing the Interest Propagation Theory}
We have assumed that interests propagate according to a diffusion-like eq. (\ref{InterestPropagationEquation}); however, the model contains free parameters ($x_{ij}$) that need to be specified. We have formulated three different hypotheses on susceptibility with increasing level of complexity that we have tested against the DBLP dataset. The parameters have been fit by the maximum likelihood outcomes.

For the sake of simplicity (and to prevent possible overfitting), we have assumed that $x_i$, $x_{ij}$, and $x_{is}$ do not depend on the specific topic $c_k$. This means that a member influences her/his neighbours with the same intensity regardless of the subject.

In general, to estimate the susceptibility parameters, we have constructed the mean square differences  $\chi^2$ between the predicted $L's$ and the observed ones:

\begin{equation}
\label{ParameterEstimation1}
\chi^2=\sum_{t, h_i, c_k} {\left[\delta L^{th} _{h_i}(c_k,t)- \delta \xi _{h_i}(c_k,t)\right]^2} = \sum_{t, h_i, c_k} {\left[ L^{th}_{h_i}(c_k.t+\Delta t) - L_{h_i} (c_k,t)  - \delta \xi _{h_i}(c_k.t)\right]^2};
\end{equation}
\noindent where  the symbol $\delta$ indicates the variation of a quantity from one year to the next.

\begin{equation}
\delta \xi (c_k,t)=\xi(c_k,t+\Delta t)-\xi(c_k,t).
\end{equation}

\noindent One performs the optimization using the $\chi^2$ as an object function, that is minimizing the deviation of prediction from observed values. 

Since the L's represent likelihoods, they must be confined to the [0, 1] range. This implies that also the $x_{ij}$ and $x_{is}$ belong to the same interval. Therefore the feasible solutions of the optimisation process must respect these constraints: 

\begin{equation}
\label{constraint}
\left\{
\begin{array}l
x_{is} \ge 0 \\
x_{ij} \ge 0 \\
\sum_j x_{ij} + x_{is}  \le 1 
\end{array}
\right.
\end{equation}

\noindent The optimum values of the parameters are achieved analytically when the point at which the gradient of the $\chi^2$ vanishes corresponds to a feasible solution:

\begin{equation}
\label{ParameterEstimation2}
\frac{\partial} {\partial \theta}  \chi^2= 0 
\end{equation}

\noindent  When the analytical solution is unfeasible, we attribute to the parameters the closest value at boundary. 

\subsection*{The \textit{Uniform Environmental Influence} Hypothesis ($HP_1$)}
The first hypothesis ($HP_1$), that we have taken into account, states that all members have the same susceptibility to trends ($x_{is} = x_{s0}$) and are not influenced by neighbours ($x_{ij} = 0$). Accordingly, the eq. (\ref{InterestPropagationEquation}) simplifies as follows:
\begin{equation}
\label{KnowlDiffModelHP0}
L_{h_i} (c_k, t+\Delta t) = (1-x_{s0})\cdot L_{h_i} (c_k, t)+ \sum_{h_j \in N_{h_i}} x_{s0} \cdot L_s(c_k,t)
\end{equation}
where, as usual, $L_{h_i}(c_k,t)$ is the probability for a person $h_i$ to be interested in the topic $c_k$ at the time t. $L_s(c_k,t)$  represents the profile of media that convey the trends.

Under this basic hypothesis, the only parameter to be estimated is $x_{s0}$ with the constraint $ 0 \le x_{s0} \le 1 $.
By introducing the \textit{deviation of the semantic profile of each author from the environment}

\begin{equation} 
\label{deviationfromenv}
\Delta^s_{h_i}  = [ L_s (c_k, t) - L_{h_i}(c_k, t) ]
\end{equation} 

\noindent the object function becomes the following:

\begin{equation}
\label{ParameterEstimationHP0-a}
\chi^2 =\sum_{t, h_i, c_k} {\left\{x_{s0} \cdot \left[ \Delta^s_{h_i} \right] - \delta \xi_{h_i}(c_k, t)\right\}^2} 
\end{equation}

\noindent where $\delta \xi_{h_i}(c_k, t)$ is the variation of the share of interest of a member $h_i$ in the topic $c_k$. Consequently, $x_{s0}$ can be calculated analitically:

\begin{equation}
\label{ParameterEstimationHP0-b}
\frac{\delta \chi^2} {\delta x_{s0}} = 0 \Rightarrow x_{s0} = \frac
{\sum_{t, h_i, c_k} \left[ \Delta^s_{h_i} \right] \cdot \delta \xi_{h_i} (c_k, t)}
{\sum_{t, h_i, c_k} \left[ \Delta^s_{h_i} \right]^2}
\end{equation}

\subsection*{The \textit{Uniform Environmental and Neighbors Influence} Hypothesis}
The second hypothesis ($HP_2$), that we have taken into account, states that all people have the same susceptibility to trends ($x_{is} = \bar x_s$) and to the neighbours ($x_{ij} = \bar x$). 

Accordingly, the eq. \ref{InterestPropagationEquation} becomes as follows:
\begin{equation}
\label{KnowlDiffModelHP2}
L_{h_i} (c_k, t+\Delta t) = (1-\bar x-\bar x_s)\cdot L_{h_i} (c_k, t)+\frac 1{|N_{h_i}|}\cdot \sum_{h_j \in N_{h_i}} \bar x \cdot L_{h_j}(c_k, t)+\bar x_s \cdot L_s(c_k, t).
\end{equation}

By introducing \textit{the average profile of the neighbours} 
\begin{equation}
\label{averageneighprofile} 
L^N_{h_i} (c_k, t) = \sum_{h_j \in N_i} \frac{1} {N_{h_i}} \cdot L_{h_j}
\end{equation}

\noindent and the \textit{deviation of the semantic profile of each author from the neighbours}:

\begin{equation}
\label{deviationfromneigh} 
\Delta^N_{h_i}  = [ L^N_{h_i} (c_k, t) - L_{h_i}(c_k, t) ]. \\
\end{equation} 

\noindent Then to estimate the parameters $\bar x_s$ and $\bar x$, the $\chi^2$ becomes:

\begin{equation}
\label{ParameterEstimationHP2-a}
\chi^2=\sum_{t, h_i, c_k} \left[ \bar x \cdot \Delta^N_{h_i} (c_k, t) + \bar x_s \cdot  \Delta^s_{h_i} (c_k, t) - \delta \xi_{h_i} (c_k, t) \right]^2
\end{equation}

The maximum likelihood equations (from general eq. \ref{ParameterEstimation2}) can be written in a compact form introducing the scalar product  

\begin{equation}
\label{scalarproduct} 
\langle \langle f \cdot g \rangle \rangle = \langle \langle f \cdot g \rangle \rangle_t = \sum_{c_k, t} f(c_k, t) g (c_k, t) 
\end{equation} %

where $ \langle \langle f \cdot g \rangle \rangle $ is the scalar product of the eq.ne \ref{scalar}, while $ \langle \cdot \rangle_t $ represents the temporal average.

Then
  
\begin{equation}
\label{hp3equation}
\left(
\begin{matrix} 
\langle \langle [ \Delta^N_{h_i}]^2 \rangle \rangle & \langle \langle \Delta^N_{h_i} \cdot \Delta^s_{h_i}  \rangle \rangle \\
\langle \langle \Delta^N_{h_i} \cdot \Delta^s_{h_i}  \rangle \rangle & \langle \langle  [\Delta^s_{h_i}]^2  \rangle \rangle \\
\end{matrix}
\right) 
\left(
\begin{matrix} 
x_i  \\
x_{is}  \\
\end{matrix}
\right)
=
\left(
\begin{matrix} 
\langle \langle  \delta \xi_{h_i} (c_k, t)  \cdot \Delta^N_{h_i} \rangle \rangle \\
\langle \langle\delta \xi_{h_i} (c_k, t)  \cdot \Delta^s_{h_i} \rangle \rangle  \\
\end{matrix}
\right)
\end{equation}
\\
where $|n_{h_i}|$ is the number of authors.

\subsection*{The \textit{Individual Environmental and Neighbors Influence} Hypothesis}
The third hypothesis ($HP_3$), that we have taken into account, states that people have both individual susceptibility to trends ($x_{is}$) and neighbours ($x_{i,j}=x_i$). 

\noindent Accordingly, the eq. \ref{InterestPropagationEquation} becomes as folllowing
\begin{equation}
\label{KnowlDiffModelHP3}
L_{h_i} (c_k,  t+\Delta t) = (1-x_i-x_{is})\cdot L_{h_i} (c_k, t)+\frac 1{|N_{h_i}|}\cdot \sum_{h_j \in N_{h_i}} x_{i} \cdot L_{h_j}(c_k, t)+x_{is} \cdot L_s(c_k, t).
\end{equation}

The object function $\chi^2$ becomes 

\begin{equation}
\label{ParameterEstimationHP3-a-deriv}
\chi^2=\sum_{t, h_i, c_k} \left[ x_i \cdot \Delta^N_{h_i} (c_k, t) + x_{is} \cdot  \Delta^s_{h_i} (c_k, t) - \delta \xi_{h_i} (c_k, t) \right]^2
\end{equation}

\noindent and the optimum solutions are given by the following equations: 

\begin{equation}
 \label{ParameterEstimationHP3-a}
\left(
\begin{matrix} 
\langle \langle [ \Delta^N_{h_i}]^2 \rangle \rangle & \langle \langle \Delta^N_{h_i} \cdot \Delta^s_{h_i}  \rangle \rangle \\
\langle \langle \Delta^N_{h_i} \cdot \Delta^s_{h_i}  \rangle \rangle & \langle \langle  [\Delta^s_{h_i}]^2  \rangle \rangle \\
\end{matrix}
\right) 
\left(
\begin{matrix} 
x_i  \\
x_{is}  \\
\end{matrix}
\right)
=
\left(
\begin{matrix} 
 \langle \langle  \delta \xi_{h_i} (c_k, t)  \cdot \Delta^N_{h_i} \rangle \rangle \\
\langle \langle\delta \xi_{h_i} (c_k, t)  \cdot \Delta^s_{h_i} \rangle \rangle  \\
\end{matrix}
\right)
\Rightarrow \mathcal{A} \cdot 
\left(
 \begin{matrix} 
x_i  \\
x_{is} \\
\end{matrix}
\right) =
\mathcal{B}
\end{equation}

Table \ref{HypothesesEvaluationOverview} presents a summary of the testing hypotheses of the \textit{interest propagation theory} and the main numerical results.

\subsection*{Numerical Results}

The results presented in Table \ref{HypothesesEvaluationOverview} show that the quality of the fit improves with the complexity of the model behind the \textit{interest propagation theory}. 
In fact, taking into account the number of degrees of freedom ($dof$) and the value of the $\chi^2 / dof$ function (representing a good index for the method), $HP_3$ fits the dataset better than $HP_2$ and $HP_1$. Optimizing $\chi^2$ resulted in some negative values of $x_i$ and $x_{is}$. In such cases, they are considered null (case $\alpha$ in Table \ref{HypothesesEvaluationOverview}). Even if $\chi^2 / dof $ increases, it is still better than that of $HP_2$. These results support the validity of the \textit{interest propagation theory}.

\begin{table}[ht]
\caption{
\bf{A Summary overview of the different hypotheses.}
}
\label{HypothesesEvaluationOverview}
\centering
\begin{tabular}{|c|c|c|c|}\hline
\textbf{Hypothesis} & \textbf{Free Parameters} & \textbf{Estimated values} & \textbf{$\chi^2 / dof$}\\
\hline
$HP_1$
& 
\(
\begin{array}c
x_{ij} = 0 \\
x_{is}  = x_{s0} 
\end{array}
\)
& 
\(
\begin{array}c
x_{ij} = 0 \\
x_{is}  = x_{s0}=0.084 
\end{array}
\)
& 
 $4.606*10^{-6}$
\\
\hline
$HP_2$
& 
\(
\begin{array}c
x_{ij} = \bar x \\
x_{is}  =  \bar x_s 
\end{array}
\)
& 
\(
\begin{array}c
x_{ij} = \bar x = 0.051 \\
x_{is}  =  \bar x_s = 0.053
\end{array}
\)
& 
$4.576*10^{-6}$
\\
\hline

$HP_3$
& 
\(\begin{array}c
x_{ij} = x_i \\
x_{is}    
\end{array}
\)
& 
\(
\begin{array}c
\bar x_ = 0.087\\

\bar x_{s} =  0.059 
\end{array}
\)
& 
$3.780*10^{-6}$
\\
\hline

$HP_{3\alpha}$
& 
\(\begin{array}c
x_{ij} = x_i > 0 \\
x_{is} > 0   
\end{array}
\)
& 
\(
\begin{array}c
\bar x_ = 0.093\\

\bar x_{s} =  0.071 
\end{array}
\)
& 
$3.920*10^{-6}$
\\
\hline
\end{tabular}

\begin{flushleft}$HP_1$: All people have the same susceptibility to trends and are not influenced by friends;

$HP_2$: All people have the same susceptibility to trends and to neighbours.

$HP_3$: People have individual susceptibility to trends and to neighbours.

$HP_{3\alpha}$: People have individual susceptibility to trends and to neighbours. In case of negative values of $x_i$ and $x_{is}$ they are considered null.

The $\chi^2$ are normalized by the degrees of freedom ($dof$) for comparison.

\end{flushleft}
\end{table}

\noindent In the next sub-section a detailed analysis of the $HP_3$ hypothesis is presented.

\subsubsection*{$HP_3$ Detailed Results}
The $HP_3$ hypothesis is more complex than the others and deserves some further discussion. 
The analysis of the best fit equations (\ref{ParameterEstimationHP3-a}) shows that there are 420290 cases where $det \mathcal{A}  \ne 0$ and 11627 cases where $det \mathcal{A}  = 0$. Hereby $\hat{x_i}$ and $\hat{x}_{is}$ will indicate the solutions of those equations when they exist.
Unfortunately in some cases those solutions are not feasible.
In the following the different cases are analyzed in details and their statistical frequencies are presented in Table \ref{DetailedAnalysis1}. 

The case I corresponds to the most common situation where the solutions $\hat{x_i}$ and $\hat{x}_{is}$ of the eq. (\ref{ParameterEstimationHP3-a}) are feasible and we can estimate members' susceptibilities to both their neighbours and the trends.

Case II represents the situation in which the solutions of the eq. \ref{ParameterEstimationHP3-a} are positive but the constraint $\hat{x_i} + \hat{x}_{is} \le 1$ is violated and
 we have attributed the closest values at the boundaries by rescaling the resulting parameters:  

\begin{equation}
\left\{
\begin{array}l
x_{i} = \frac{\hat{x}_{i}}{\hat{x}_{i} + \hat{x}_{is}} \\
x_{is} = \frac{\hat{x}_{is}}{\hat{x}_{i} + \hat{x}_{is}}.
\end{array}
\right.
\end{equation}

\begin{figure}[!ht]
\begin{center}
\includegraphics[width=0.85\textwidth, angle=0]{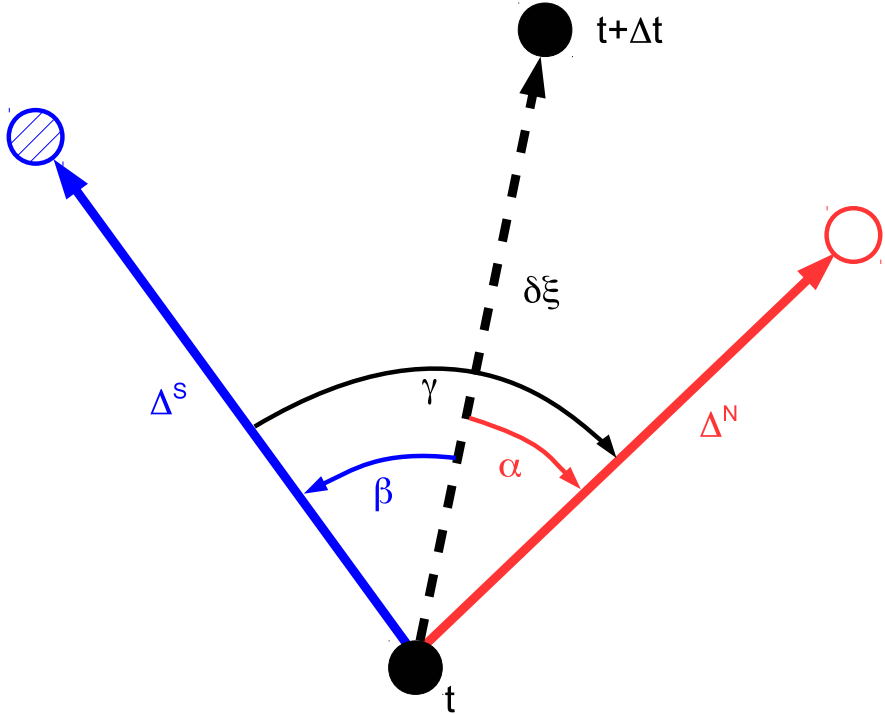}
\end{center}
\caption{Pictorial representation of one step of the interest diffusion. Each point represents a possible profile. The black filled dots represent the profile for a member at two subsequent times ($t$ and $t+\Delta t$); the red empty dot represents the average profile of her/his neighbours; and the blue dashed dot represents the profile of the total trends source. The solution of the eq. \ref{hp3equation} corresponds to the linear combination of $\Delta^N$ and $\Delta^S$ closest to $\delta \xi$.} 
\label{angles}
\end{figure}

In case III, $\hat{x_i}$ is negative while $\hat{x}_{is}$ is feasible. From the point of view of the diffusion equation this means that the member is not influenced by the neighbours but by the trends only and we have to attribute a null value to $x_{i}$.  However, there can be other interpretations.
A negative value of $x_i$ indicates that the change in the profile $\delta \xi_{h_i}$ has a component directed against $\Delta^N_{h_i}$. From the geometrical point of view (see figure \ref{angles}), this corresponds to the trigonometric inequality: $cos ( \alpha ) < cos (\beta) \cdot cos (\gamma)$.

\noindent There can exist members that tend to publish in different subjects with respect to their neighbours. This can take place when the members want to exhibit some kind of independency with respect to other people in the group they belong to. 
This type of behaviour is not included in our present model. Accounting for such an effect would result in a non diffusive dynamics.
 
Negative values may also be a spurious consequence of the overlapping of the different multi-lexemes representing the topics. 
In fact, if there are similarities or synonyms in the basic set of topics, small angles $\alpha$ between members and their neighbours can appear close to $\pi$.
Again we stress that the quality of the semantic analysis plays the most important role. 
 
In case IV $\hat{x_i}$ is positive while $\hat{x}_{is}$ is negative. As in the previous case the interpretation from the diffusion equation is that the member is not influenced by the trends. 
However also in this case there are alternative interpretations. The most important effect that is neglected in the theory is innovation. 
Some members introduce new topics. This results into some negative $cos(\beta)$. Current limits of the semantic analysis may also result into spurious large $\beta$.     

Case V corresponds to negative values of both $\hat{x_i}$ and $\hat{x}_{is}$. The same considerations of case III and case IV apply here.

In case VI, $det \mathcal{A}  = 0$, the matrix is singular.  $x_i$ can not be determined.

\begin{table}[ht]
\caption{
\bf{HP3: Individual features estimation}}
\label{DetailedAnalysis1}
\centering
\begin{tabular}{|c|c|c|c|c|c|}\hline
\textbf{$Case$} & \textbf{$detA$} & \textbf{$x_i$} & \textbf{$x_{is}$} & \textbf{$x_i + x_{is}$} & \textbf{$Occurrences$}\\
\hline
\hline
Ia
&
$det \mathcal{A} \not= 0$ 
&
$\hat{x}_i =x_i > 0$
& 
$\hat{x}_{is} = x_{is} > 0$
& 
$x_i + x_{is} \le 1$
& 
$191239$
\\
\hline

Ib
&
$det \mathcal{A} \not= 0$ 
&
$\hat{x}_i =x_i = 0$
& 
$\hat{x}_{is} =x_{is} = 0$
& 
$x_i + x_{is} < 1$
& 
 $117473$
\\
\hline

Ic
&
$det \mathcal{A} \not= 0$ 
&
$\hat{x}_i =x_i > 0$
& 
$\hat{x}_{is} =x_{is} = 0$
& 
$x_i + x_{is} \le 1$
& 
$3$
\\
\hline
\hline

II
&
$det \mathcal{A} \not= 0$ 
&
$\hat{x}_i \ne x_i > 0$
& 
$\hat{x}_{is} \ne x_{is} > 0$
& 
$\hat{x}_i  + \hat{x}_{is} > 1$
& 
$462$
\\
\hline
\hline

III
&
$det \mathcal{A} \not= 0$ 
&
$\hat{x}_i < 0 \to x_i = 0$
& 
$0<x_{is} \le 1$
& 
$x_i + x_{is} \le 1$
& 
$54694$
\\
\hline
\hline

IVa
&
$det \mathcal{A} \not= 0$ 
&
$\hat{x}_i =x_i > 0$
& 
$\hat{x}_{is} < 0 \to x_{is} = 0$
& 
$x_i + x_{is} \le 1$
& 
$55621$
\\
\hline

IVb
&
$det \mathcal{A} \not= 0$ 
&
$\hat{x}_i > 1 \to x_i = 1$
& 
$\hat{x}_{is} < 0 \to x_{is} = 0$
& 
$\hat{x}_i+ \hat{x}_{is} > 1$
& 
$463$
\\
\hline
\hline

V
&
$det \mathcal{A} \not= 0$ 
&
$\hat{x}_i < 0 \to x_i = 0 $
& 
$\hat{x}_{is} < 0 \to x_{is} = 0 $
& 
$x_i + x_{is} < 1$
& 
$335$
\\
\hline
\hline

VIa
&
$det \mathcal{A} = 0$ 
&
$\hat{x}_i$ is undetermined 
& 
$0 < \hat{x}_{is}=x_{is} < 1$
& 
$ - $
& 
$3125$
\\
\hline

VIb
&
$det \mathcal{A} = 0$ 
&
$\hat{x}_i$ is undetermined 
& 
$\hat{x}_{is}=x_{is} = 0$
& 
$-$
& 
$8440$
\\
\hline

VIc
&
$det \mathcal{A} = 0$ 
&
$\hat{x}_i$ is undetermined 
& 
$\hat{x}_{is} < 0 \to x_{is} = 0$
& 
$-$
& 
$107$
\\
\hline

\end{tabular}
\begin{flushleft}
$det \mathcal{A}$ is the discriminant of the eq. (\ref{ParameterEstimationHP3-a}), $\hat{x_i}$ and $\hat{x}_{si}$ are their solutions and $x_i$ and $x_{is}$ are the estimated features. 
Case I: Analytic best fit solutions are feasible. 
Case II: Members are positively influenced by both the trends and the neighbours but the analytic best fit solutions are unfeasible.
Case III:  Members are positively influenced by the trends but the analytic best fit solution $\hat{x_i}$ is negative.
Case IV:  Members are positively influenced by the neighbours but the analytic best fit solution $\hat{x}_{is}$ is negative.
Case V: The analytic best fit solutions $\hat{x}_i$ and $\hat{x}_{is}$ are negative.
Case VI: Members susceptibility by their neighbours are undetermined. 
\end{flushleft}
\end{table}
 
\begin{figure}[!h]
\begin{center}
\includegraphics[width=1.00\textwidth, angle=0]{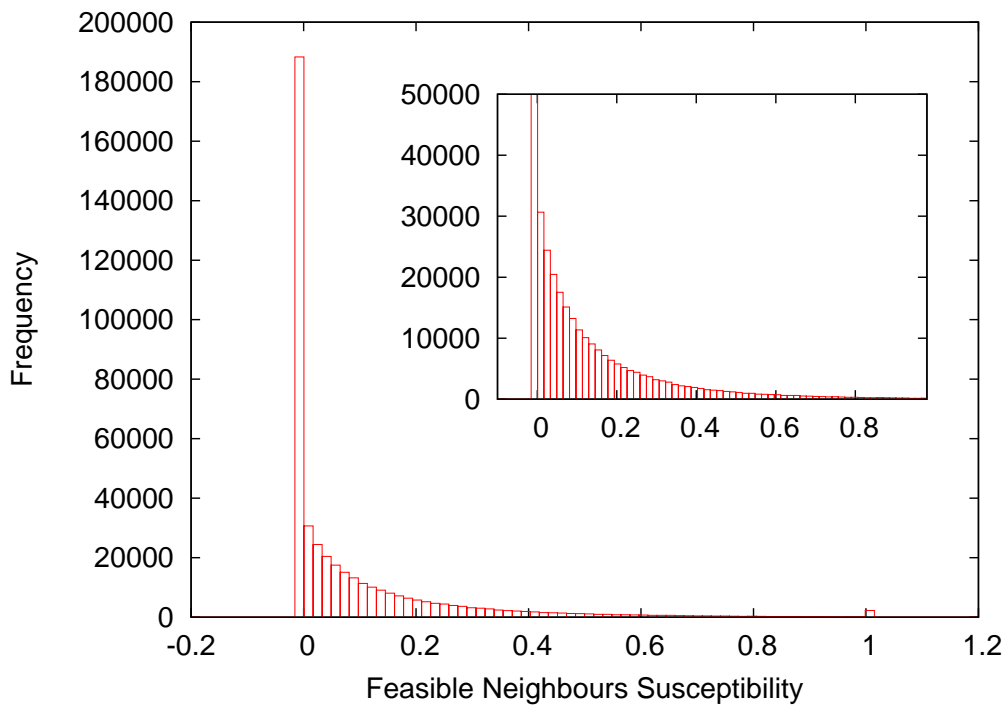}
\end{center}
\caption{Histogram of neighbours' fitted susceptibilities (feasible solutions only).}
\label{fig-suscnei-feasible}
\end{figure}

\begin{figure}[!h]
\begin{center}
\includegraphics[width=1.00\textwidth, angle=0]{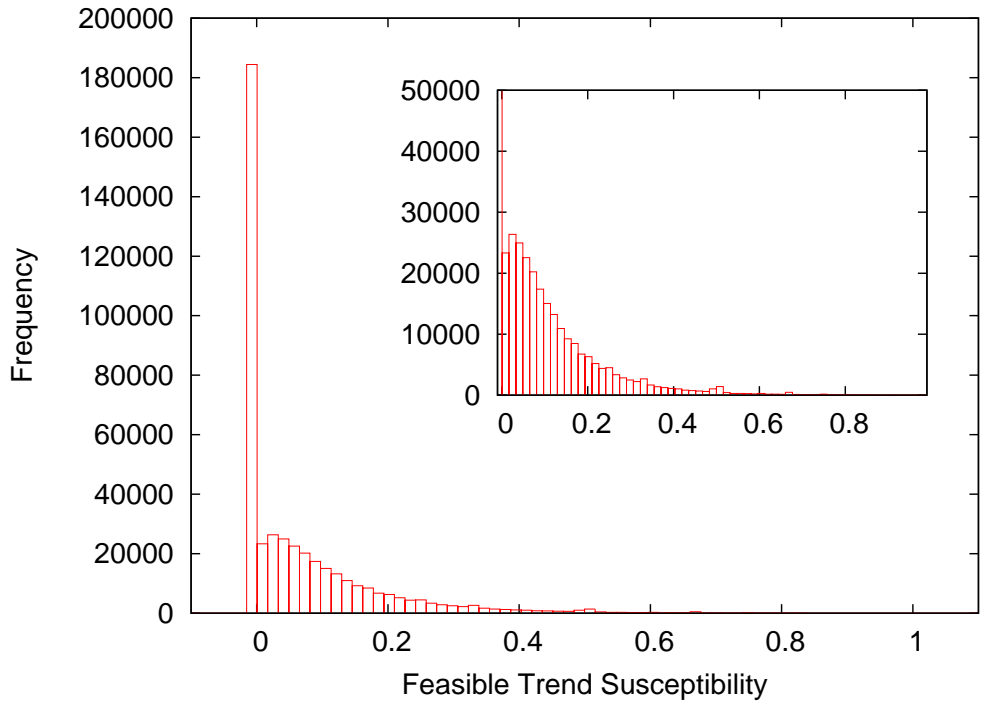}
\end{center}
\caption{Histogram of trend's fitted susceptibilities  (feasible solutions only)}
\label{fig-suscmoda-feasible}
\end{figure}

The distributions of susceptibilities compatible with the constraints of the diffusion theory are reported in the figures \ref{fig-suscnei-feasible} and \ref{fig-suscmoda-feasible}.  
The average susceptibility under the $HP_{3\alpha}$ hypothesis due to neighbours is $9.3 \%$, whereas the contribution due to trends is $7.1 \%$ for a total average susceptibility of $16.4\%$.
Roughly speaking, this means that about $85\%$ of the subjects of publications are along the line of the previous works while some 15\% do exhibit new topics due to the influence of collaborators and trends.
The distribution profiles show a very pronounced peak at null susceptibility, while being smooth for other values.
The existence of such peak may result from genuine effect, but it may also be an artifact of an insufficient semantic analysis.
As a matter of fact, there is a large set of papers (1215200 out of 2246098) that can not be indexed by means of our selected set of topics. This is expected to result into spurious null susceptibilities. In order to test whether this is the case we have created a second (wider) set of topics imposing the constraint that all papers must be indexed. The resulting new set consists of 120917 topics. All papers were indexed; however, due to its size, the semantic analysis of this new set was not accurate enough to prevent undesired synonyms, mutually similar and even fake topics (the latter were, in fact, eliminated from the set of 7632 topics by human inspection). The computational time necessary to test the new wider set against the whole database is prohibitive for our present computational capabilities, therefore, we have limited the analysis to a reduced period of time (from 1985 to 1990).
\begin{table}[ht]
\caption{Some statistics of trends ($x_{is}$) and neighbours' ($x_i$ ) susceptibilities employing two different sets of indexing topics (time period: 1985-1990)}
\label{1985-1990Susc}
\centering
\begin{tabular}{|c|c|c|c|c|}
\hline
\textbf{Indexing Set Size} & \textbf{$x_i <0$} & \textbf{$x_i=0$} & \textbf{$0 \le x_{i} \le 1$} & \textbf{$x_i >1$}\\
\hline
\hline

7632 
&
$15.01\%$
& 
$43.59\%$
& 
$41.05\%$
& 
$0.35\%$
\\
\hline

120917
&
$21.60\%$
& 
$13.43\%$
& 
$64.46\%$
& 
$0.51\%$
\\
\hline
\hline
\textbf{Indexing Set Size} & \textbf{$x_{is} <0$} & \textbf{$x_{is}=0$} & \textbf{$0 \le x_{is} \le 1$} & \textbf{$x_{is} >1$}\\

\hline

7632 
&
$4.40\%$
& 
$38.10\%$
& 
$57.49\%$
& 
$0\%$
\\
\hline

120917
&
$3.19\%$
& 
$3.14\%$
& 
$93.67\%$
& 
$0\%$.
\\
\hline

\end{tabular}
\end{table}

\noindent As can be seen from Table \ref{1985-1990Susc} the wider sample is capable to reduce significantly the spurious null neighbours susceptibilities, while increasing the number of negatives. Such an increasing  is worth attributing to the semantic redundancy and quality of the extracted topics. More precisely, similarities tend to provide extra topics to publications while they are actually on the line of previous work with old coauthors. Concerning the susceptibility to trends, the extension of the set of basic topics results in a significant decrease of both the number of null values (as expected) and negatives.  This is consistent with the previous interpretation, as similar topics are all represented in the trends and switching from one to another can not be read as contrasting the trends. 

It is evident that the semantic analysis represents one of the most critical points in our work. Further analyses (that are subject of ongoing work) will possibly provide a better set of basic topics with minimum similarity and maximum coverage.

\begin{figure}[!h]
\begin{center}
\includegraphics[width=1.00\textwidth, angle=0]{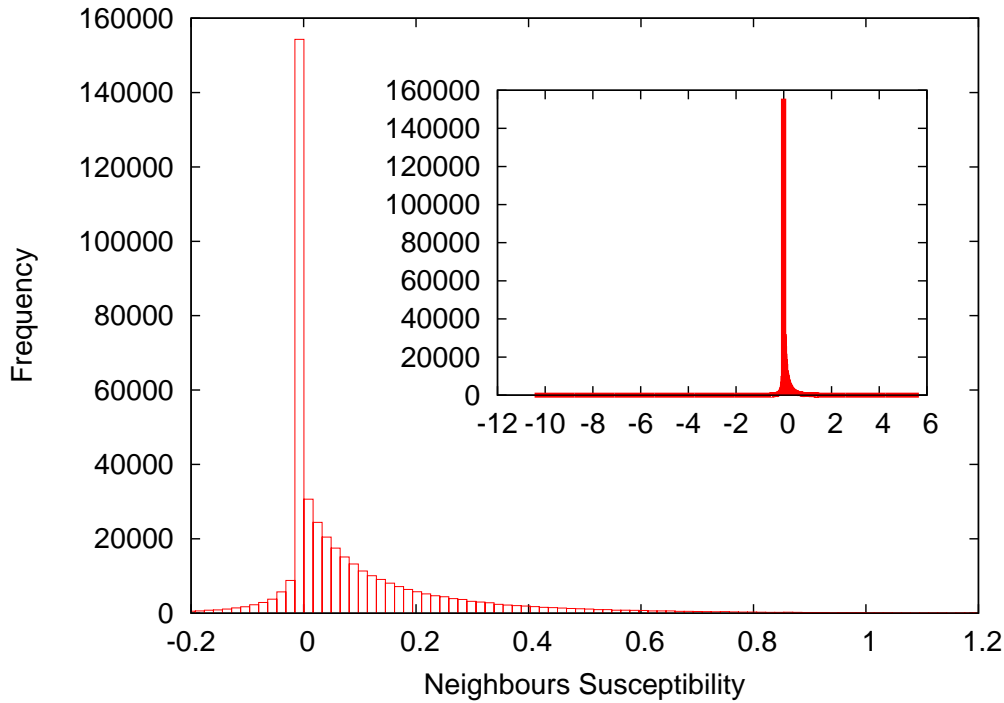}
\end{center}
\caption{Histogram of susceptibilities to neighbours. Feasibility constaints are released.}
\label{fig-neisusc-nobound}
\end{figure}

\begin{figure}[!h]
\begin{center}
\includegraphics[width=1.00\textwidth, angle=0]{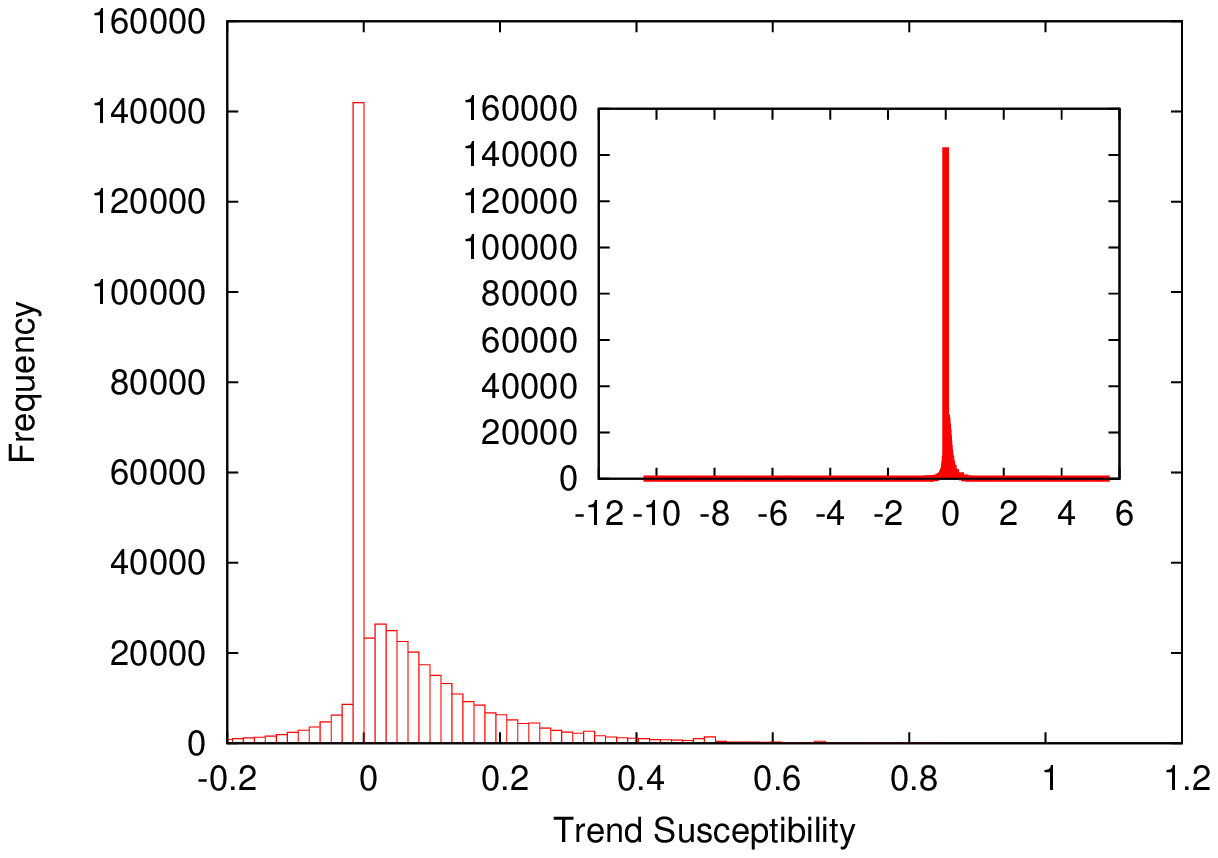}
\end{center}
\caption{Histogram of susceptibilities to trends. Feasibility constaints are released.}
\label{fig-suscmoda}
\end{figure}

The distributions of susceptibilities achieved disregarding the constraints of the theory ($HP_3$) are reported in the figures \ref{fig-neisusc-nobound} and \ref{fig-suscmoda}. In this case, the average susceptibility to neighbours is 8.7\%, while the contribution from trends is 5.9\%, for a total average susceptibility of 14.6\%. As a variation from the feasible solutions (Figs  \ref{fig-suscnei-feasible} and \ref{fig-suscmoda-feasible}), one observes a significant decrease of null values of about $ 13\%$ in both cases, corresponding to the negative solutions. 

\begin{figure}[!h]
\begin{center}
\includegraphics[width=1.00\textwidth, angle=0]{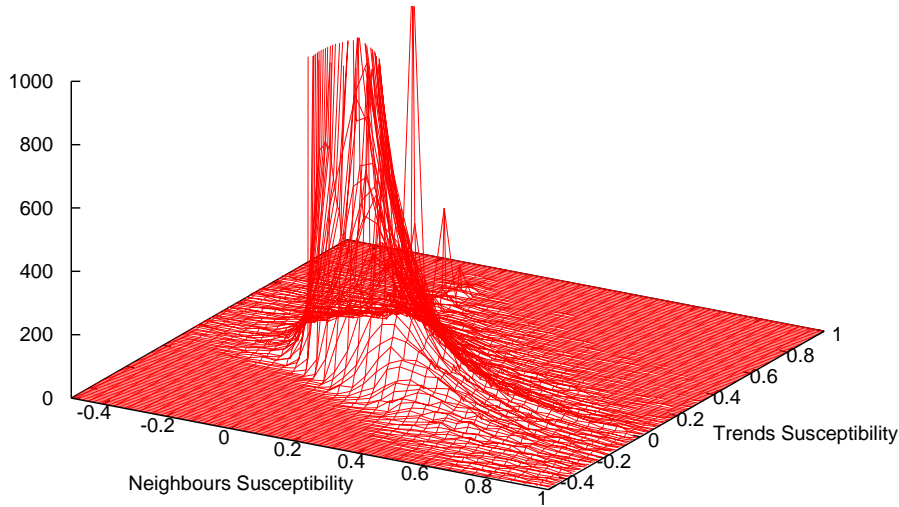}
\end{center}
\caption{Three dimensional histogram of the frequencies of the fitted $hat{x}_i$ and $\hat{x}_{is}$. The majority of the values lay in the feasible region, yet significant negatives are observed.}
\label{3Dxixsihistogram}
\end{figure}

Generally speaking, under the $HP_3$ hypothesis, the susceptibility to neighbours is higher than that to trends ($\bar x > \bar x_{s}$). However the two susceptibilities are anti-correlated (with a correlation coefficient of about $r = - 0.4$); that is the total susceptibility fluctuates less then each (neighbours of trends) component. Figure \ref{3Dxixsihistogram} shows the distribution of $x_i$ and $x_{is}$ frequencies as a three dimensional histogram. Most of the values lay in the feasible range ($ 0 \le x_i \le 1 $ and 
$ 0 \le x_{is} \le 1 $), yet several points lay outside. Moreover, consistently with the observed anti-correlation, there is a significant concentration along the bisector of the second and forth  quarters. This general considerations are stable against the variation of the set of basic topics, while details may be strongly influenced by some ambiguities in the dataset (see afterwords) and by the quality of semantic analysis.

\begin{figure}[!h]
\begin{center}
\includegraphics[width=1.00\textwidth, angle=0]{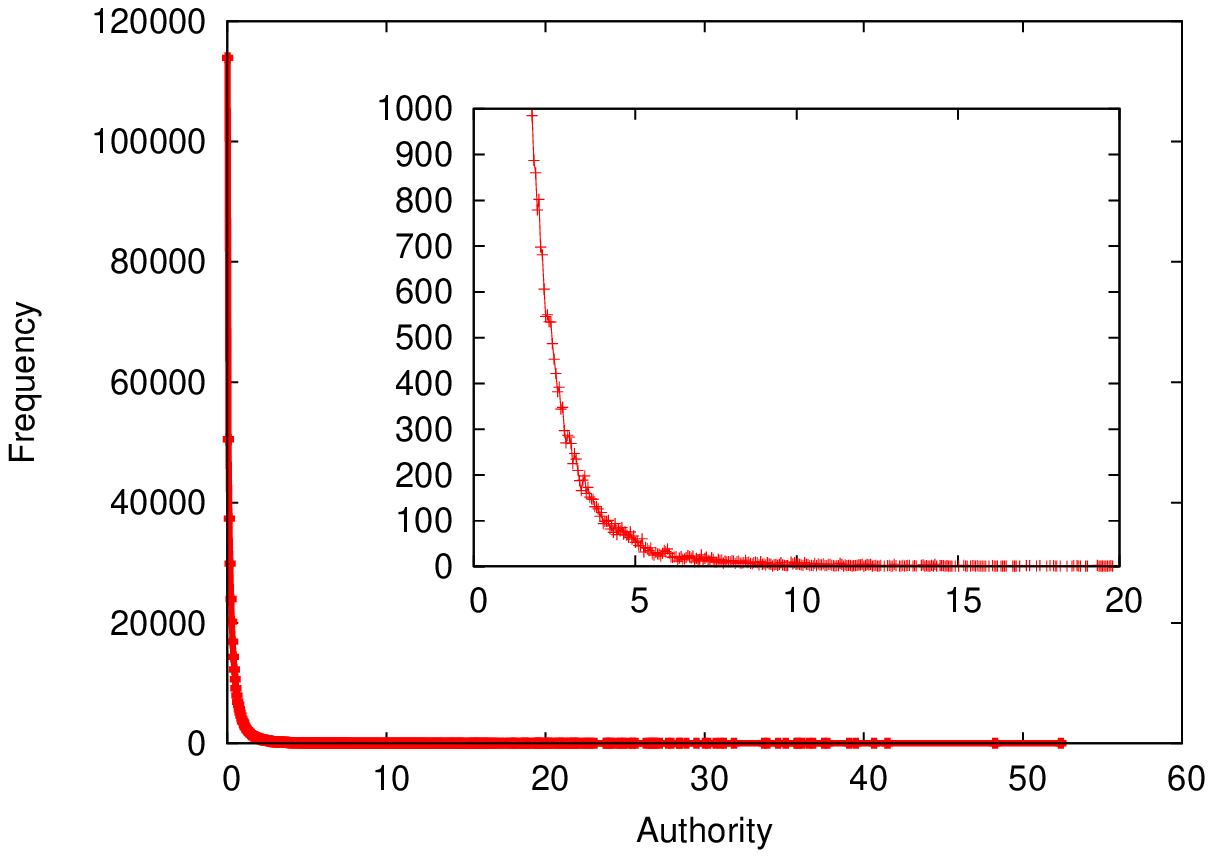}
\end{center}
\caption{Histogram of the authority coefficients.}
\label{AuthorityCoeffHist}
\end{figure}

\begin{figure}[!h]
\begin{center}
\includegraphics[width=90mm]{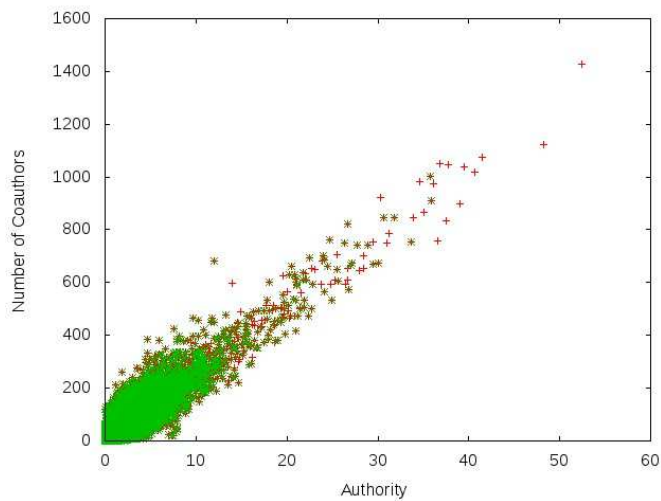}
\end{center}
\caption{Scatter plot with the relationship between the number of the coauthors of an author and her / his authority. Scatter plot with the relationship between the number of the coauthors of an author and her / his authority filtering the Chinese most frequent names.  Red crosses represent Chinese authors with a frequent full name whereas green crosses represent other authors.}
\label{FigCoauthVSauthorityT}
\end{figure}

The authority coefficients (authorities, shortly) spans the [0,52] range; their mean value is $\bar a = 0.44$ and its standard deviation is about twice that value (0.89). 
Figure \ref{AuthorityCoeffHist} shows the histogram of authority distribution. As can be seen, there is a very long queue of few authors at high values. 
While there certainly exist real authors with some hundred collaborators, some of the observed ones may be fictitious.
It is known that there exist different authors with the same name (given name and family name); 
those people are very often treated as a single author in several datasets. 
This problem is known as "ambiguity" of the papers indexing; it results in gathering different authors into a single member of our social network. 
Due to the many to one transliteration of the original names in latin characters, Asian members are mostly prone to such effect.
The problem is known to affect DBLP data analysis \cite{Han:2010:MKD:1807167.1807333}. 
In order to check this phenomenon, we have built a list of frequent Chinese full names by combining 50 very common Chinese given names \cite{wiki:chinesenames} with 100 frequent Chinese family names \cite{wiki:chinesesurnames}.
Figure \ref{AuthorityCoeffHist} shows a scatter plot of the  authorities as a function of the number of publications. 
Several high values of authority correspond to entries associated with the constructed set of frequent Chinese full names. 
As it is clear from figure \ref{FigCoauthVSauthorityT}, the relationship between the authorities and the number of coauthors is not linear. 

Table \ref{FamousAuthors} presents some individual features of some famous authors. As expected they all exhibit high levels of authority.

\begin{table}[ht]
\caption{Famous authors in Computer Science}
\label{FamousAuthors}
\centering
\begin{tabular}{|c|c|c|c|}\hline
\textbf{Name} & \textbf{$\hat{x}_i$} & \textbf{$\hat{x}_{is}$} & \textbf{Authority $a_i$}\\
\hline
\hline

\textbf{Wil M. P. van der Aalst} 
&
$+0.111$
& 
$+0.058$
& 
$+12.809$
\\
\hline

\textbf{Jack Dongarra} 
&
$-0.019$
& 
$+0.028$
& 
$+10.259$
\\
\hline

\textbf{John Mylopoulos} 
&
$+0.021$
& 
$+0.037$
& 
$+8.852$
\\
\hline

\textbf{Georg Gottlob} 
&
$+0.055$
& 
$+0.009$
& 
$+5.081$
\\
\hline

\textbf{Ian Horrocks} 
&
$+0.198$
& 
$-0.080$
& 
$+4.835$
\\
\hline

\textbf{Maurizio Lenzerini} 
&
$+0.106$
& 
$-0.065$
& 
$+3.128$
\\
\hline

\textbf{Erol Gelenbe} 
&
$+0.184$
& 
$+0.015$
& 
$+3.123$
\\
\hline

\end{tabular}
\end{table}

\begin{figure}[!h]
\begin{center}
\includegraphics[width=1.00\textwidth, angle=0]{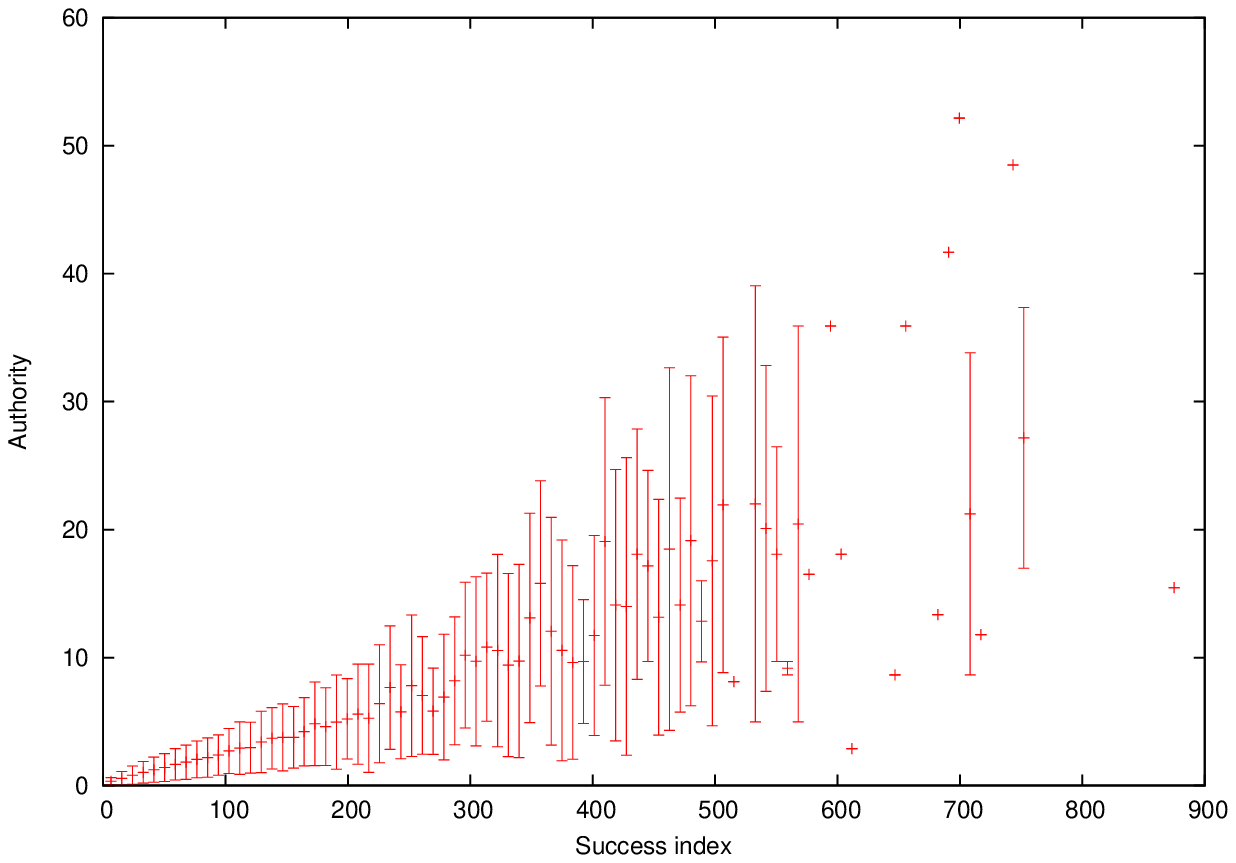}
\end{center}
\caption{Scatter plot of the authority of different semantically treatable authors versus their success index.}
\label{AuthvsSucc}
\end{figure}

\begin{figure}[!h]
\begin{center}
\includegraphics[width=1.00\textwidth, angle=0]{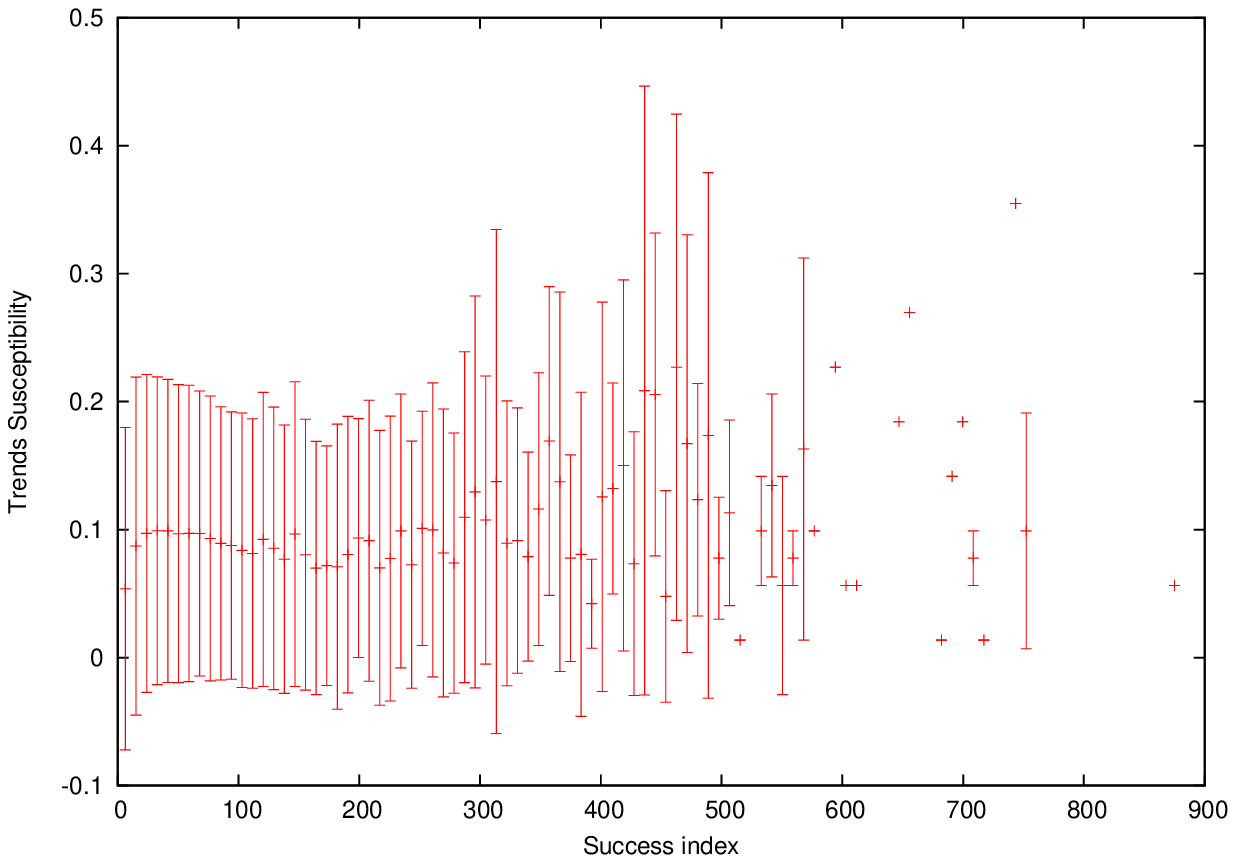}
\end{center}
\caption{Scatter plot of trends susceptibilities versus the success index.}
\label{TrendvsSucc}
\end{figure}

We have also tried to relate the success of an author with its authority. We have employed the number of published papers as an index of success, however a more appropriate index should be the total number of citations \cite{Petersen28102014, Wang04102013} which where not available.  As shown in Figure \ref{AuthvsSucc}, the higher is the success index the higher is the authority. 
Figure \ref{TrendvsSucc} presents the relationship between the success index and the trends susceptibility. In this case, there is not a clear dependency between the two variables as the same values of the success index correspond to different values of trends susceptibility. This means that there are successful authors of different types: some of them follow trends; some propose new topics and some continue working mostly on the same topics. 

\section*{Discussion}

This paper represents a first attempt to combine known methods in complexity science with the semantic analysis of natural language to provide insights in the propagation of people interests in social networks. 
As  a first basic model we have assumed that interests propagate according to a diffusion mechanism, while being continuously created. 
The human behaviour is described by means of two basic behavioural characteristics (the susceptibility and the authority)
 that quantify the tendency to influence and being influenced by "friends" and the environment.

The original ideas were developed having in mind the social networks and especially their instantiations on internet platforms. However the theory applies also to scientific co-authorship networks. Since data for this type of networks were easily available we have tested the consequences of our model on one of them, the DBLP, that gathers a wide set of papers published in computer science. 

The preliminary results are very encouraging and, in fact, the diffusion mechanism, seems to be a leading part of the story.  A single thread does propagate (like gossips) according to epidemic equations 
(\ref{InterestPropagationEquation}). However, interests in a field are more persistent in people and tend to be shared among "friends", thus leading to the diffusion mechanism. 

In this preliminary version of the model a lot of relevant factors have been purposely neglected. Among several others, it is worth stating some of them explicitly: 

 The ageing of the links \cite{zhu2003effect,tutoky2013weights}. In our model, once a link is established it is supposed to hold forever; whereas in reality links can also vanish for several reasons related to competitiveness, displacements, personal frictions etc. In general, we do not take into account the strengths of links and their evolution. 
 
 The multiplex problem \cite{mucha2010community,gomez2013diffusion,granell2013dynamical}. People have different types of relationships and, hence, there exist multiple networks that may convey their interests; whereas our present model acquires information only from one source. This problem may also be framed in the contest of the hidden links research.
 
The Semantic profiling \cite{middleton2004ontological,sheth2001system}. Generally speaking one is not allowed to assume that there exist a set of disjoint topics covering all possible interests. Concepts are normally overlapping and possibly one interest may induce an other one in a "close" topic. Typical examples come from music genders. To attribute a person a specific interest in a gender such as jazz and not to blues is rather unreasonable. In the present model, in order to provide members with a  semantic profile,  the existence of a basic set of disjoint topics is strictly required. Further developing will enter the semantic structure of the Domain of Interest and will lead to more complex modelling.

Psychological types. In our model, we only allow people to be influenced by friends (or by the environment) or to be independent on them. 
However, there are several reasons for which a person may deliberately decide to do things, not just disregarding friends' positions, but in contrast with them. 
This maybe for competition or just for spirit of independence. This type of behaviour is usually referred to as "anti conformity"  and it has been studied elsewhere\cite{nyczka2013anticonformity,willis1965conformity,pronin2007alone}.

Beside the limits of the theory there are other factors that need further treatment and have possibly hindered the analysis of the DBLP: the semantic analysis and the disambiguation.

Even assuming that there exist a basic set of disjoint topics covering the Domain of Interest, the semantic analysis may only lead to an approximation of it. Small sets of basic topics are not capable to index all papers, while larger ones tend to contain synonyms, similar multi-lexemes and, above all, concepts not representing interests (e.g. the words: report, surveys etc). As already discussed in the Section "$HP_3$ Detailed Results", in our experimentation an effect of the lack of coverage of the domain is the presence of null values of $x_i$ and $x_{is}$. 

The member identification in the DBLP suffers from ambiguities. There are author names under which papers by different members are gathered (polysemy) and, viceversa, there are authors that sign different papers with slightly different names (synonymy).  This issue is particularly relevant for Asian names. 
The problem is currently approached by different authors with promising results \cite{ferreira2012brief}, \cite{wang2014unified}.

The second issue concerns the quality and the completeness \cite{burton2005semiotic} of the identified topics . 

The third issue concerns semantic profile estimation. Currently, interests are extracted from the titles of the papers by means of natural language techniques. 
A suitable wider set of sources will allow to detect a more complete set of topics and, hence, to define more accurate profiles. 

Despite the limits of the present paper, it demonstrates that diffusion of interest on social networks is a reality and it can be exploited to provide 'ad hoc' services to their members.

\section*{Acknowledgments}

The authors kindly acknowledge fruitful discussions with Antonio Scala and Dmitry Zinoviev.

%
%

%
%
%

%
%

%

%
%

%

%

\end{document}